\documentclass[prd,preprint,superscriptaddress]{revtex4}
\usepackage{amsmath}
\usepackage{amssymb}
\usepackage{revsymb}
\usepackage{graphicx}
\usepackage{epstopdf}

\usepackage{array}
\usepackage{float}
\usepackage{tabularx}
\usepackage{booktabs}

\newcommand{\be}{\begin{equation}}
\newcommand{\ee}{\end{equation}}
\newcommand{\ben}{\begin{eqnarray}}
\newcommand{\een}{\end{eqnarray}}

\newcommand{\lsim}{\lower.7ex\hbox{$\;\stackrel{\textstyle<}{\sim}\;$}}
\newcommand{\gsim}{\lower.7ex\hbox{$\;\stackrel{\textstyle>}{\sim}\;$}}

\begin{document}

\title{Primordial gravitational waves spectrum in the interacting Bose-Einstein gas model}
\author{Germ\'an Izquierdo}
\address{Facultad de Ciencias, Universidad Aut\'onoma del Estado de M\'exico, Toluca 5000, Instituto literario 100, Estado de  M\'exico, M\'exico }
\email{gizquierdos@uaemex.mx}
\author{Gildardo Alonzo}
\address{Facultad de Ciencias, Universidad Aut\'onoma del Estado de M\'exico, Toluca 5000, Instituto literario 100, Estado de  M\'exico, M\'exico }
\author{Jaime Besprosvany }
\address{Instituto de F\'{\i}sica, Universidad Nacional Aut\'onoma de M\'exico, Apartado Postal 20-364, Ciudad de M\'exico 01000,   M\'exico}
\date{\today}
\begin{abstract}
We study the evolution and power spectrum of primordial gravitational waves in the interactive Bose-Einstein gas  model for dark energy, relevant, as it  addresses the coincidence problem. The model is applied in the radiation, matter and dark-energy domination stages.
 The model introduces  a  scale factor associated to the radiation-matter transition which influences  the gravitational spectrum.
We focus on the impact of the free parameters    on both the gravitational waves amplitude and its power-spectrum slope. For sets of parameters   fitting  Hubble's law, we show that   the model's parameter for today's dark-matter energy density
 has a noticeable impact on such waves, while the others produce an indistinguishable  effect.  The feasibility of detecting such waves under  present and future  measurements is discussed.
\end{abstract}

\maketitle

\section{Introduction}
Gravitational waves (GW) are second-order tensorial propagating wave-like solutions to General Relativity's field equations that were predicted by Einstein back in 1916 \cite{Einstein1916}. GW are generated by different kinds  of sources (pulsars, merging Black Holes, the Big Bang) and their evolution equations can be obtained by considering them as a perturbation of a corresponding background metric. Indirect observation of the GW through the period variation of the binary pulsar PSR1913+16 was obtained in 1975  \cite{Hulse75}. More recently, LIGO and VIRGO succeeded in observing GW directly \cite{ligo16}.

Cosmology can  take advantage of   the GW physics in different ways. Astrophysical
events that emit GW as well as electromagnetic  radiation can be used to estimate the  Universe   expansion rate    (Hubble constant) \cite{schutz86, ligo17}.  This estimation is particularly useful to solve the problem of the Hubble-constant tension
 between its measurements  made by the  PLANCK Project \cite{ade2016planck} and by the Hubble Space Telescope (HST) \cite{Riess20162}. Primordial Gravitational Waves (PGW) are perturbations of the Friedman-Lemaitre-Robertson-Walker  metric whose amplitude evolves during the    universe expansion in a   characteristic way \cite{Lifs, grish74}.   PGW  have a small energy density on the ground-based detector  frequencies
which make very difficult the direct detection. In fact, the PGW  power spectrum   can be bounded by  the Laser Interferometer Gravitational-Wave
Observatory (LIGO) experiment \cite{stochLIGO17}. Future space-based detectors like eLISA, the Laser Interferometer Space Antenna, \cite{eLISA} could change this scenario and open the window to direct detection. Also, B-mode polarization on the Cosmic Microwave Background radiation is generated by low-frequency PGW  present at the last-scattering surface \cite{sel}. It would be possible to reconstruct or at least to put bounds on the PGW  spectrum through
 this kind of observed data \cite{cheng}.

The Universe is experiencing a late-stage  accelerated expansion
 \cite{ade2016planck} induced by an unknown  energy-density  source  called
 Dark Energy (DE). Although the cosmological constant model is the most favoured by observational data \cite{ade2016planck}, it is plausible to consider other DE models \cite{copeland2006dynamics}.
  In particular, coupled DE models describe a dark-sector interaction (i.e. DE interacts with cold dark matter through a coupling term)
\cite{Wang2016dmdeinteract},  addressing the coincidence problem.

Exchange models are constructed based on   physical  processes. In a class of model-dependence  on the number-density and energy-density components,  for the  IBEG model term, it emerges from a fluid decay process common in astrophysical processes  \cite{barrow2006}; other models rely on the dependence on the energy-density components' time derivative \cite{Shahalam15}, with a description of stable fixed points. In these models, attractor solutions  solve the coincidence problem. Dark-energy evolution can also  be ascribed to a time dependence of the cosmological component \cite{SRay11}, as such a model emerges from scalar-field models at the inflationary time \cite{Dymnikova00,Dymnikova01}

While the early-expansion model (inflation and phase transitions)
   is fundamental to the   PGW amplitude evolution, the late stage has also an important impact on the low-frequency wave amplitude and power spectrum \cite{iz04}. In \cite{sasaki18}, the authors find an exact solution for PGW in a universe with a cosmological constant. In \cite{Almazan14}, some coupled DE models are considered as well, describing the amplitude and power spectrum of the PGW in terms of the model's free parameters.

The Interacting Bose-Einstein Gas (IBEG)  model   assumes the DE is a gas of non-relativistic Bose-Einstein self-interacting particles; for the  late-expansion description it couples to cold dark matter (CDM) in a way that the IBEG particle number changes  with the expansion \cite{iz10, bespro2015}.
The IBEG model has a detailed microscopic description and the model's free parameters can be bounded by observational data \cite{Luc18}.  In this work, we study the evolution and power spectrum of PGW in the IBEG model. We focus on the impact of the
 free-parameter  choices on both the PGW amplitude and its power-spectrum slope.
 We demonstrate how the choice of the parameter $\Omega_{m0}$, related to CDM mass density energy, has a noticeable impact on the PGW, while the rest
of the  parameters lead to a similar amplitude and power spectrum. Obtaining observational data of the low-frequency PGW power spectrum could help   bound parameter $\Omega_{m0}$ of the IBEG model.

The plan of the article is as follows: In section \ref{s2}, we briefly review the PGW  amplitude evolution equations. In section \ref{s3}, we
address the IBEG universe dynamics and compute
 PGW  amplitudes for different    free-parameter choices. In section \ref{s4}, we estimate the power spectrum of the PGW. Finally, in section \ref{s6}, we summarize the findings.

We assume units for which $c =\hbar=k_B=1$. As usual, a zero subindex refers to the current value of the corresponding quantity; likewise, we normalize the scale factor of the metric by setting $a_{0} = 1$.

\section{PGW evolution from the Big Bang until the radiation era}
\label{s2}

We define $h_{\alpha \beta}$ as perturbations of the background Lemaitre-Friedman-Robertson-Walker (LFRW) metric. The total metric reads $\overline{g}_{\alpha \beta}=g_{\alpha \beta}+h_{\alpha \beta}$,
$\left\vert h_{\alpha \beta}\right\vert \ll \left\vert g_{\alpha \beta}\right\vert $, $\alpha, \beta=0,1,2,3$. The background $g_{\alpha \beta}$ is the flat homogeneous and isotropic LFRW metric
\[
ds^{2}=-dt^{2}+a(t)^{2}\left[ dr^{2}+r^{2}d\Omega ^{2}\right] =a(\eta
)^{2}[-d\eta ^{2}+dr^{2}+r^{2}d\Omega ^{2}],
\]%
where $t$ and $\eta $ are, respectively, the cosmic and conformal time ($a(\eta )d\eta ={dt}$), with comoving coordinates,
$r$, the   radius, and $\Omega$,  the solid angle.

To obtain  the  sourceless perturbation evolution   to linear
order (PGW), we choose the transverse-traceless tensor gauge. The resulting
equations can be expressed as \cite{Lifs, grish74}%
\[
h_{i j}(\eta ,{\mathbf{x}})=\int h_{i j}^{({\mathbf{k}})}(\eta ,{\mathbf{x}}%
)d^{3}{ k },
\]%
\begin{equation}
h_{i j}^{({\mathbf{k}})}(\eta ,{\mathbf{x}})=\frac{\mu (\eta )}{a(\eta )}%
G_{i j}({\mathbf{k}},{\mathbf{x}}),
\end{equation}%
where space indices use latin letters and run from $1$ to $3$,  $ {\mathbf{x}}$ is the comoving Cartesian coordinate, and  ${\mathbf{k}} $ is the comoving wave vector. The functions $G_{i j}({\mathbf{k}},{\mathbf{x}})$  satisfy the equations
\begin{equation}
{G_{i }^{j}}_{;m }^{;m }=-k^{2}G_{i }^{j},\qquad {G_{i }^{j}}_{;j}=G_{i }^{i }=0,  \label{G}
\end{equation}%
implying for  $\mu (\eta )$

\begin{equation}
\mu ^{\prime \prime }(\eta )+\left [ k^{2}-\frac{a^{\prime \prime }(\eta )}{%
a(\eta )}\right ] \mu (\eta )=0,  \label{eqmu}
\end{equation}%
where the prime indicates derivative with respect to the  conformal time,
and $ k=\left\vert{\mathbf{k}}\right\vert $ is the constant wave number, related to the
physical wavelength  and frequency by $k=2\pi a/\lambda =2\pi af=a\ \omega$. The functions ${G_{i }^{j}}$ are combinations of $\exp (\pm i%
\mathbf{{k}}\cdot\mathbf{{x}})$, which contain the two possible wave polarizations,
compatible with the conditions (\ref{G}).

Eq. (\ref{eqmu}) is a time-independent Schr\"{o}dinger
equation with potential term $a^{\prime \prime }/a$. When $k^{2}\gg \frac{a^{\prime
\prime }}{a}$, i.e., for waves whose wavelength is smaller than the horizon, expression (\ref{eqmu})
becomes a free-wave equation.
The    $h_{i j}^{({ \mathbf {k} })}(\eta ,{\mathbf{x}})$ amplitude tends to null
adiabatically as $a^{-1}$ in an expanding universe. In the opposite regime,
when $k^{2}\ll \frac{a^{\prime \prime }}{a}$, i.e., when the PGW wavelength
is larger than the horizon, the solution to (\ref{eqmu})
is a lineal combination of $\mu _{1}\propto a(\eta )$ {and }$\mu _{2}\propto
a(\eta )\int d\eta \ a^{-2}$. In an expanding universe, $\mu _{1}$ grows
faster than $\mu _{2}$ and will soon dominate. The
  $h_{i j}^{({ \mathbf{k}})}(\eta ,{\mathbf{x}})$  amplitude is constant while  the  condition $k^{2}\ll \frac{a^{\prime \prime }}{a}$ is fulfilled. When the PGW reenter the
 horizon, the wave will have an amplitude greater than it
would in the adiabatic behavior. This phenomenon is known as \textquotedblleft superadiabatic
amplification\textquotedblright\ of PGW \cite{grish93,
grish74}.

For sources with constant equation of state, the resulting scale factor is a  power-law expansion $a\propto \eta ^{l}$ ($l=-1,1,2$ for de Sitter,
radiation dominated and dust-dominated universes, respectively). Equation (\ref{eqmu}) is a Bessel equation with solution
\[
\mu (\eta )=(k\eta )^{\frac{1}{2}}\left[ C_{1}J_{l-\frac{1}{2}}(k\eta
)+C_{2}J_{-\left( l-\frac{1}{2}\right) }(k\eta )\right ] ,
\]%
where $J_{l-\frac{1}{2}}(k\eta )$, $J_{-\left( l-\frac{1}{2}\right) }(k\eta
)$ are Bessel functions of the first kind and $C_{1,2}$ are integration
constants.

We assume now that the early universe experiences an inflationary de Sitter stage of evolution, followed by a radiation-dominated stage, and a dust stage
\cite{grish93}. Transitions
between successive eras are assumed instantaneous. This approach is known as the sudden transition approximation,
which  is   reasonable  when the transition time span between the different stages is much lower than the period of the PGW  considered. The scale factor, then, is%
\begin{equation}
a(\eta )=\left\{
\begin{array}{c}
-\frac{1}{H_{1}\eta }\qquad\ \ \ \ \   -\infty <\eta <\eta _{1}<0 , \\
\frac{1}{H_{1}\eta _{1}^{2}}(\eta -2\eta _{1})\qquad \ \ \ \ \  \eta _{1}<\eta <\eta
_{2}  , \\
\frac{1}{4H_{1}\eta _{1}^{2}}\frac{(\eta +\eta _{2}-4\eta _{1})^{2}}{\eta
_{2}-2\eta _{1}}{\qquad }\ \ \ \ \  \eta _{2}<\eta ,  %
\end{array}%
\right.  \label{sclfac1}
\end{equation}%
where the subindexes $1,2$ correspond to the sudden transitions from
the inflation to the radiation era and  from the radiation to the  dust era,
respectively, and $H_{1}$
represents the Hubble factor at the end of the inflationary era.
The solution to  eq. (\ref{eqmu}) for each era is

\begin{eqnarray}
\mu _{I}(\eta ) &=&C_{I}\left[ \cos (k\eta +\phi _{I})-\frac{1}{k\eta }\sin
(k\eta +\phi _{I})\right] \qquad \text{(inflationary era)}  \label{muinfl} \\
\mu _{R}(\eta ) &=&C_{R}\sin (k\eta _{R}+\phi _{r})\qquad \qquad \qquad
\qquad \qquad \qquad \text{(radiation era)} \label{murad}\\
\mu _{D}(\eta ) &=&C_{D}\left[ \cos (k\eta _{D}+\phi _{D})-\frac{1}{k\eta
_{D}}\sin (k\eta _{D}+\phi _{D})\right] \qquad \text{(dust era)},
\label{mudust}
\end{eqnarray}
where $C_{I,R,D}$, $\phi _{I,R,D}$ are  integration  constants, $\eta
_{R}=\eta -2\eta _{1}$ and $\eta _{D}=\eta +\eta _{2}-4\eta _{1}$ .

It is possible to express $C_{R}$, $\phi _{R}$ \ and $C_{D}$, $\phi _{D}$ in
terms of $C_{I}$, $\phi _{I}$ as $\mu(\eta )$ must be continuous at the transition times $\eta =\eta _{1}$ and $%
\eta =\eta _{2}$. Averaging the solution over the initial phase $\phi _{I}$, the amplification factor is

\begin{eqnarray}
\frac{C_{D}}{C_{I}}\sim \left\{
\begin{array}{c} \!\!\!\!\!\!\!\!\!\!
1 \qquad \qquad  k\gg -1/\eta _{1} , \\
  k^{-2} \qquad -1/\eta _{1}\gg k\gg 1/ \eta _{D2} , \\ \!\!\!\!\!\!\!\!\!\!\!\!\!\!\!
k^{-3}\qquad \qquad 1/ \eta _{D2} \gg k ,%
\end{array}%
\right.
\end{eqnarray}
where $\eta _{D2}=2\eta _{2}-4\eta _{1}$.

The PGW  evolution from the  dust era up to the present day depends on the
late-acceleration stage considered. As the universe experiences  such a stage, the potential term $a''/a$ becomes an increasing
function of $\eta$. Consequently,  some waves that were already  in the $k^2\gg a''/a$ regime reenter
the $k^2\ll a''/a$ regime, and cease contibuting to the PGW  physical power spectrum.   In \cite{iz04}, the amplification is computed for different constant equations of state in  dark-energy models. In \cite{Almazan14}, two coupled
dark-energy scenarios are considered. In \cite{sasaki18}, an exact solution to the late acceleration ruled by the cosmological constant is found. In all cases, it is important  that the late-stage acceleration universe   leaves a
 characteristic amplification on low-frequency waves.

In the next section, we consider a different scenario in which a coupled IBEG stage follows the radiation era.

\section{PGW evolution from the radiation era up to present time}
\label{s3}
\subsection{IBEG model and expansion factor}
The IBEG  model for the late-acceleration stage assumes the universe has three energy-density sources: baryonic matter $\rho_{b}$, cold dark matter (CDM) $\rho_{dm}$ and the IBEG
$\rho_{g}$ \cite{bespro2015,Luc18}. The latter is a gas of
Bose-Einstein
particles that self-interact attractively with non-null kinetic energy.  An energy flux is imposed between
the IBEG and the CDM, which  induces the non-condensate  IBEG particle
number density  to evolve   as $n_{\epsilon}=n_{\epsilon 0}a^{3(x-1)}$, where $n_{\epsilon 0}$ is the IBEG number density today, and $x$ is the parameter that models the Markoff variation process of the IBEG particles\footnote{For a particle Markoffian creation process the number of particles evolves as $N \propto V^{x}$ where $V$ is the volume considered and $x$ is a parameter as the distribution of future states of it does not depend on previous
states. Typical dispersion processes and fluid interactions lead to evolution laws of this
type in various physical setups \cite{bespro2015}.}. In \cite{bespro2015}, the parameter $x$ is found to be in the range $0.85\leq x\leq 1$. On one hand, $x\geq 1$ would lead to an IBEG number density increasing with expansion, which eventually would make the IBEG energy density negative as well, while for $x<0.85$, the IBEG model does not solve the coincidence problem.  The gas energy density and pressure evolve with the expansion as \cite{bespro2015}
\ben
\rho_g &=& \rho_{G0} a^{3(x-1)}+ \rho_{c0} a^{5(x-1)}+\rho_{i0} a^{6(x-1)},\label{rhogpg}\\
p_g&=&\frac{2}{3}\rho_{c0} a^{5(x-1)}+\rho_{i0} a^{6(x-1)},\label{pg}
\een
where $\rho_{G0}$ is the model's free parameter connected the IBEG particles' mass, $\rho_{c0}$ relates to the IBEG kinetic energy, and $\rho_{i0}$ is the self-interaction term ($\rho_{i0}<0$).

As our IBEG model relies on  an attractive self-interaction, we address
the model's gas  stability. In \cite{Khlopov85}, the gravitational stability  of a scalar field is studied with a repulsive/attractive self-interaction $\mp\lambda^2 \phi^4$ in  the Newtonian approximation. The authors find that the Jeans instability is similar to that of a dust model with extra `hydrodynamic'  self-interaction effects.
In particular, for the attractive case ($+\lambda^2 \phi^4$), the obtained  Jeans wave number is larger than that of the corresponding free scalar field. In the IBEG case, we model a phenomenological short-range two-particle attractive   potential through a contact interaction \cite{iz10,bespro2015}, so naturally, $+\lambda^2 \phi^4$ is a valid quantum description in the low-density limit. We conclude that the IBEG instabilities would evolve in a similar way to instabilities in \cite{Khlopov85}. Also, in \cite{bespro2015}, the   linear density-perturbation evolution  and corresponding equations were obtained in the cosmological expanding background for both the coupled IBEG gas and the CDM, taking into account the coupling term.

In \cite{Dymnikova00,Dymnikova01}, a Bose condensate models the early-universe accelerated
expansion. The Higgs field considered self-interacts through the  potential $V(\phi)=\frac{\lambda}{4} (\phi^2-\phi_0^2)^2$. Such an interaction naturally accounts for slow-roll inflation as well as for the reheating process, as real particles are created   by  the   Higgs particles'  decay.
In \cite{iz10}, a non-coupled IBEG gas containing both condensate and non-condensate particles  is considered in an early universe approach, solving the horizon problem with a
super-exponential expansion for some parameter  choices. Such a growth allows for density-fluctuation propagation for a large range of scales, suggesting that they  are scale invariant, and that primordial fluctuations can be generated.

Addressing the coincidence problem  for the universe late-expansion description, some models include  coupling between CDM and Bose-Einstein particles.
In \cite{SRay11}, the Bose-Einstein condensate constitutes  a time-varying   $\Lambda$ term while the   condensate decay (with the same mechanism described in \cite{Dymnikova00,Dymnikova01}) produces a  coupling proportional to $\dot{\Lambda}$. This approach has differences with the IBEG model. For the latter model, it is the CDM that decays on the IBEG particles (for most parameters), while  for the model in \cite{SRay11}, the opposite is the case. In addition, we investigated the case in which the Bose-Einstein particles are created in a  non-condensate state, as we concentrate on the  non-null kinetic-energy term.


The energy density evolution equations for the IBEG model read
\begin{eqnarray}
\dot{\rho}_{b}\, &+&\, 3H \rho_{b} = 0 \, , \nonumber \\
\dot{\rho}_{dm}\, &+& \, 3 H \rho_{dm} = -Q \, , \nonumber \\
\dot{\rho}_{g}\, &+& \, 3 H (\rho_{g}+p_g) =  Q ,
\label{continuity}
\end{eqnarray}
where $H=\dot{a}/a$ is the Hubble expansion factor and $Q$ is the coupling term. From eqs. (\ref{rhogpg}-\ref{pg}) and the  above equations, one  obtains the coupling term
\be Q=3Hx \left(\rho_{G0}\,{a}^{3x-3}+\frac{5}{3} \rho_{c0}a^{5x-5}+2\rho_{i0}a^{6x-6} \right). \label{Q}   \ee
Some coupling terms   in the literature are proportional to the Hubble factor $H$ from dimensional analysis considerations, since $H$ is the characteristic  time-inverse FLRW quantity \cite{Wang2016dmdeinteract}. Other coupled models assume heuristically that $Q$ is proportional to the  DE and/or CDM energy-density time derivatives,   not directly depending on the Hubble factor  \cite{Shahalam15}. In our model, the coupling (\ref{Q}) is proportional to the Hubble factor, which derives from the dependence of $n_{\epsilon}$ on the scale factor and the IBEG microscopic description.

The CDM energy density    is solved as
\be\rho_{dm} =\rho_{m0}a^{-3}-\rho_{G0}a^{3(x-1)}+\frac{5x\rho_{c0}}{2-5x}a^{5(x-1)}+\frac{2x\rho_{i0}}{1-2x}a^{6(x-1)},\label{rhodm}\ee
where $\rho_{m0}$ is an integration constant representing the CDM  energy density  due to its mass today.

Given that the baryonic matter evolves as $\rho_{b}=\rho_{b0}a^{-3}$ ($\rho_{b0}$
 is the baryonic-matter energy density today), the Hubble factor satisfies \cite{Luc18}

\be
H^2=\frac{8 \pi G}{3} ( \rho_{b}+\rho_{dm}+\rho_{g}) =H_0^2\left[(\Omega_{b0}+\Omega_{m0})a^{-3}+\frac{2\Omega_{c0}}{2-5x}a^{5(x-1)}+\frac {\Omega_{i0}}{1-2x}a^{6(x-1)}\right],
\label{H}
\ee
where $\Omega_{a0}= 8 \pi G \rho_{a0} /(3H_0^2)$ with $a=b,m,c,i,$ and $H_0$ is the present-day Hubble expansion rate.  The present day DM energy density is defined from (\ref{rhodm}) as
\be
\Omega_{dm0}=\Omega_{m0}-\Omega_{G0}+\frac{5x\Omega_{c0}}{2-5x}+\frac{2x\Omega_{i0}}{1-2x}.
\ee
We  observe that  $\Omega_{m0}$ is not the present-day energy density of DM, but only the term that evolves as $a^{-3}$; aside from it, $\Omega_{dm0}$    depends the other parameters $\Omega_{i0}$, $\Omega_{G0}$, $x$   related to the coupling between DM and IBEG particles.

The  IBEG-model free parameters are $H_0, \Omega_{m0},
\Omega_{i0}, \Omega_{b0}, x$ and $\Omega_{G0}$, the latter not appearing in the Hubble factor.  $\Omega_{c0}$ is related to these parameters, as we assume a flat LFRW metric

\be
\Omega_{c0}=\frac{2-5x}{2}\left(1-\Omega_{b0}-\Omega_{m0}-\frac {\Omega_{i0}}{1-2x}\right).
\label{Omc}
\ee
Given the  IBEG microscopic description with parameter $\Omega_{i0}<0$ evolving with the scale factor as $a^{6(x-1)}$, the IBEG energy density $\rho_g$ would tend in the past to negative values when $x\ne1$ \cite{bespro2015}. We can avoid this problem by assuming the flux of energy from CDM to IBEG is an ongoing process that starts no sooner than the instant for which $\rho_g=0$. We consider then the scale factor $a_{in}$ defined as the solution to
\be \rho_g( a_{in})= \rho_{G0} a_{in}^{3(x-1)}+ \rho_{c0} a_{in}^{5(x-1)}+\rho_{i0} a_{in}^{6(x-1)}=0 ,\ee
for a given  free-parameter set, as the instant at which the creation of IBEG particles starts. The scale factor $a_{in}$, on its own, is not a new parameter of the model but it is dependent on the free parameters considered. The creation process starting at  $a_{in}$ represents   a natural solution under the assumptions made for the gas (both the microscopic description of the gas and the particle creation rate) \cite{bespro2015}. The coupling between CDM and IBEG particles    is common in    astrophysical processes\cite{barrow2006}, giving support to the  interaction between particles that leads to the creation rate assumed and that starts at  $a_{in}$.   On the other hand, for $x= 1$, $\rho_g$ is constant and does not tend to null at early stages of evolution. In this case, there is no need to consider $a_{in}$.

\begin{figure}
\includegraphics[scale=0.39]{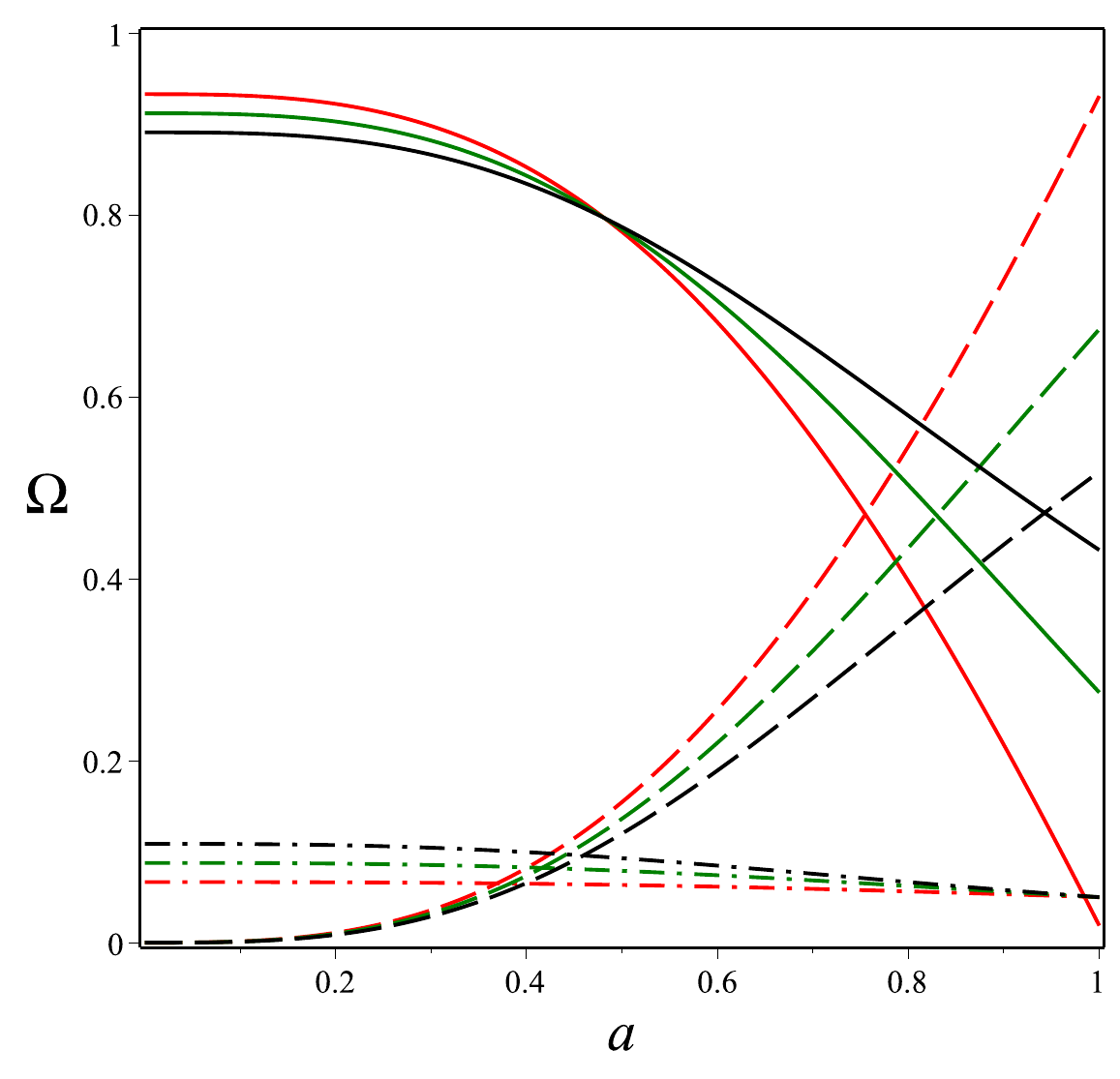}
\includegraphics[scale=0.4]{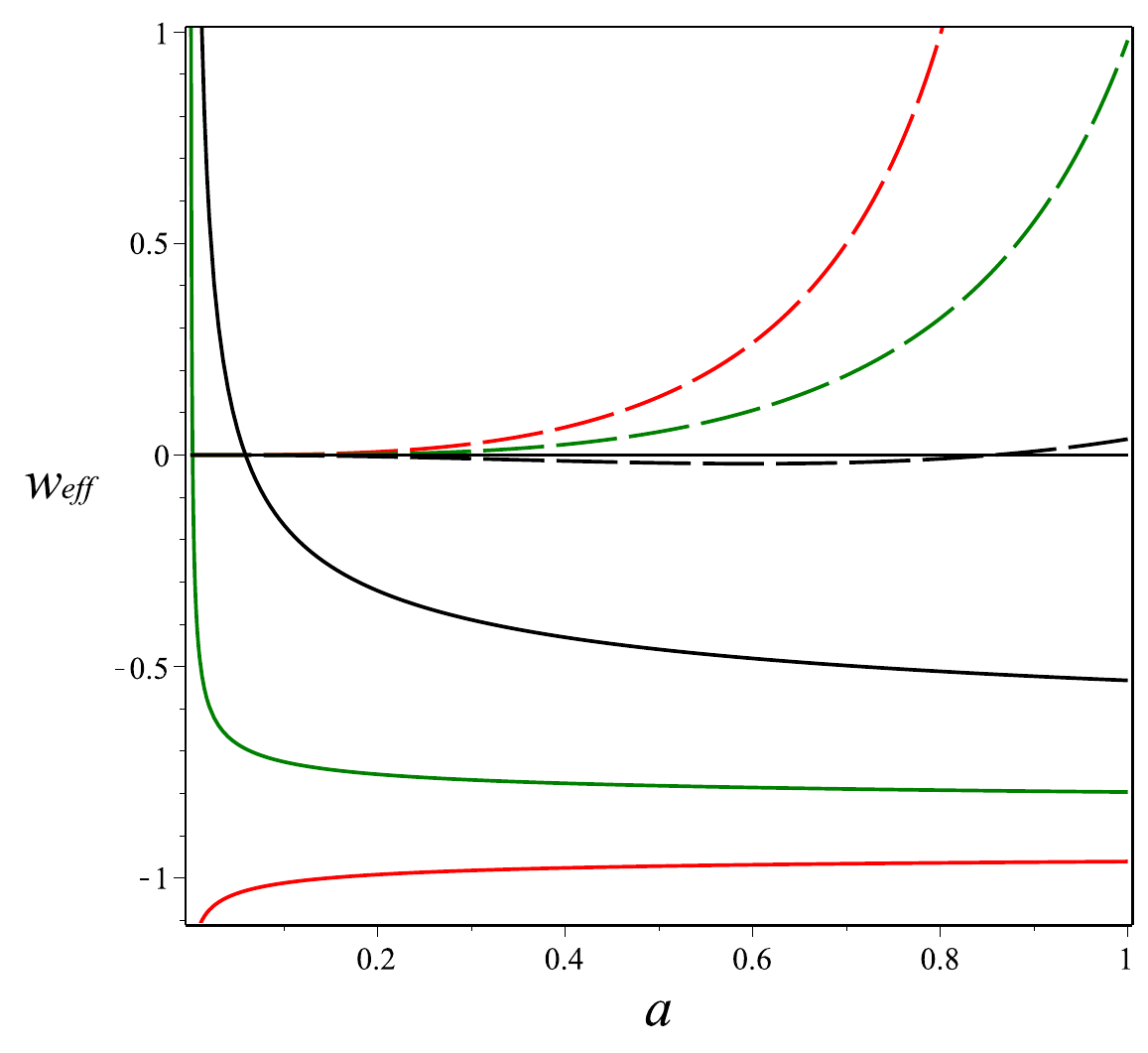}
\caption{Left panel: Evolution of $\Omega_{dm}$ (solid lines), $\Omega_{g}$ (dashed lines) and $\Omega_{b}$ (dot-dashed lines) vs. scale factor $a$ for $x=0.97$, $\Omega_{b0}=0.05$, $\Omega_{i0}=-1.1$, $\Omega_{G0}=0.72$ and three choices of parameter $\Omega_{m0}$: $0.4$ (black lines), $0.52$ (green lines), and $0.7$ (red lines). Right panel: Evolution of effective adiabatic parameters $w_{eff}^{(g)}$ (solid lines) and $w_{eff}^{(m)}$ (dashed lines) vs. the scale factor $a$ for the same choice of parameters as in the left panel.}
\label{fig0}
\end{figure}

The left panel of figure \ref{fig0} shows the evolution of $\Omega_{dm}=3\rho_m/(8\pi G H^2)$, $\Omega_{g}=3\rho_g/(8\pi G H^2)$ and $\Omega_{b}=3\rho_m/(8\pi G H^2)$ vs. scale factor $a$ for different choices of the free parameters. The right panel of figure \ref{fig0} shows the evolution of effective adiabatic parameter for both CDM ($w_{eff}^{(m)}=Q/\rho_m$) and the effective adiabatic parameter of the IBEG fluid ($w_{eff}^{(g)}=(p_g-Q)/\rho_g$) vs. scale factor $a$ for different choices of free parameters. The effective adiabatic parameter is often considered in coupled dark-energy models and can be defined from the energy by moving the coupling term to the left-hand side of the corresponding conservation equation.

\begin{figure}
\centering
\includegraphics[scale=0.5]{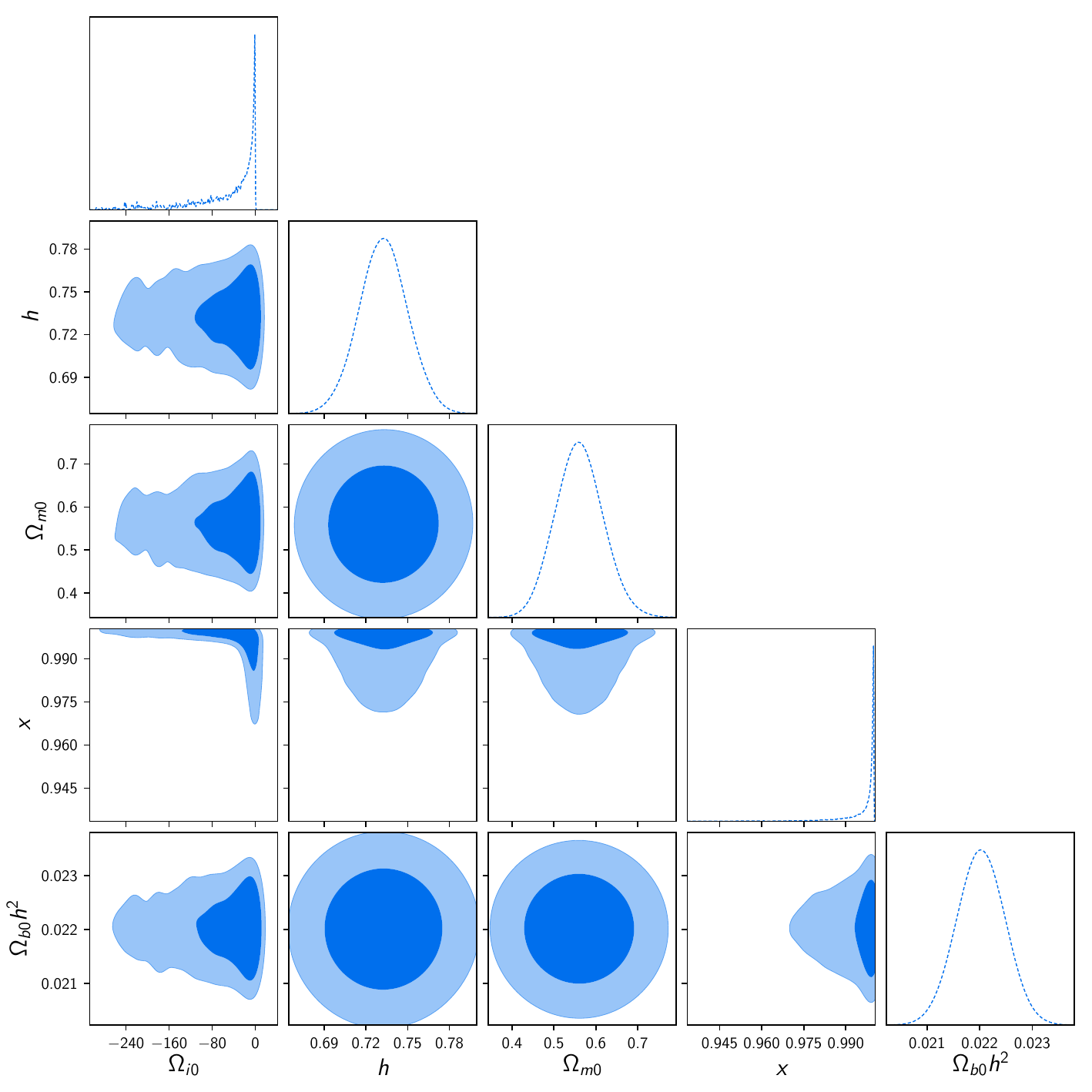}
\includegraphics[scale=0.5]{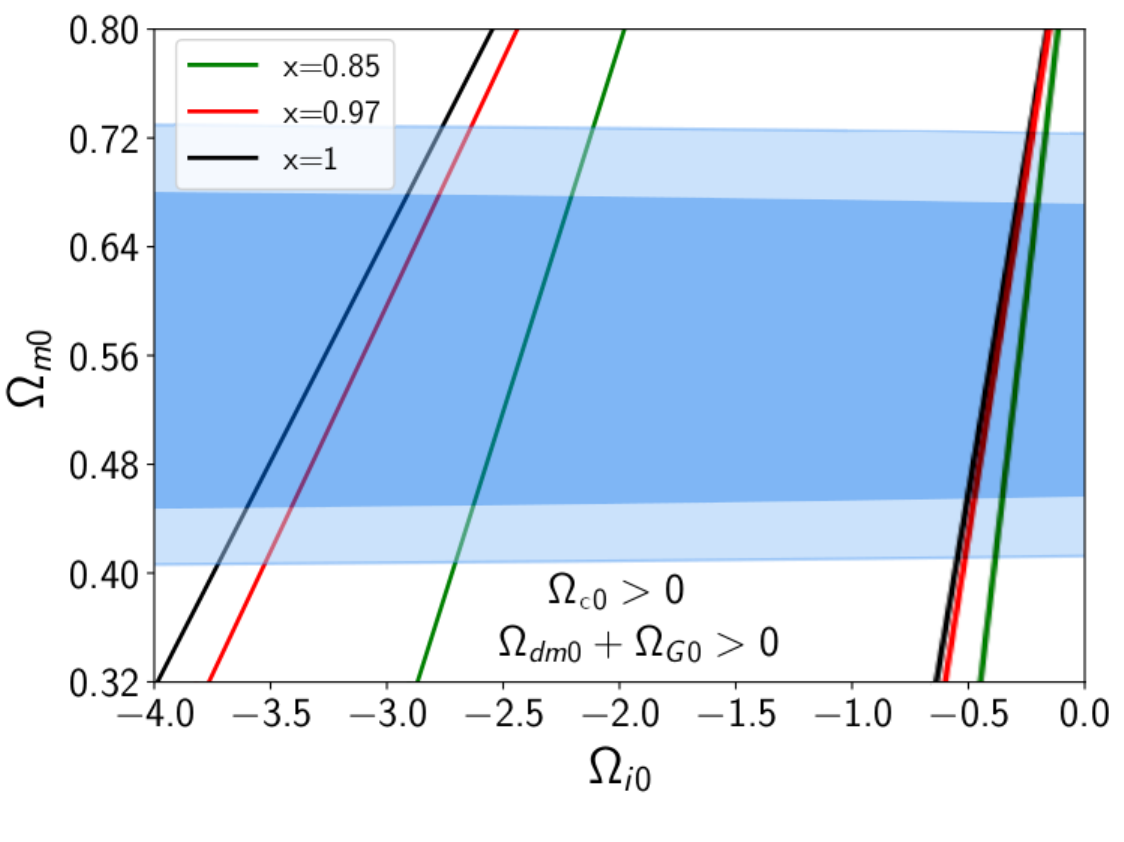}
\caption{Top panel:  The $1\sigma$ and $2\sigma$ likelihoods for the free parameters as obtained in \cite{Luc18}. Bottom panel: The $1\sigma$ and $2\sigma$ likelihoods for the free parameters in the $\Omega_{m0}$ \, vs \, $\Omega_{i0}$ space for $\Omega_{b0}H_0^2=0.022$, $H_0=70\  {\rm  km} /({\rm Mpsc}\,{\rm s})$ shown as obtained in \cite{Luc18}. Lines on the left side of the plot represent the bound $\Omega_{dm0}+\Omega_{G0}=0$ given by (\ref{thcond}) for three different choices of $x$: $x=0.85$ (green line), $x=0.97$ (red line), and, $x=1$ (black line). Lines on the right side represent the bound $\Omega_{c0}=0$. The space closed by the lines represents the parameter choices with $\Omega_{dm0}+\Omega_{G0}>0$ and $\Omega_{c0}>0$.}
\label{fig1}
\end{figure}

In Ref. \cite{Luc18}, the expansion rate  in eq. (\ref{H}) is used to adjust the free parameters to three independent sets of Hubble-factor observational data. The best-fit values for the free parameters obtained with the corresponding $1\sigma$ likelihood are $H_0=70\pm2\ { \rm km} /({\rm Mpsc}\, {\rm s})$, $\Omega_{m0}=0.52\pm0.08$, $\Omega_{i0}=-3.60\pm12.38$,
$\Omega_{b0}H_0^2=0.022\pm0.001$ and $x=097\pm0.01$. The results are shown in figure \ref{fig1}, together with two additional theoretical bounds. The first one  emerges from  the CDM     particle mass and IBEG component,  which is positive definite
\ben
\Omega_{dm0}+\Omega_{G0}
 = \frac{5x}{2}-\frac{5x}{2}\Omega_{b0}+\frac{2-5x}{2}\Omega_{m0}-\frac{x}{2(1-2x)}\Omega_{i0}\geq 0.\label{thcond}
\een
This bound is represented by the lines on the figure's lhs    for different $x$. The second bound is  $\Omega_{c0} > 0$,   represented by the lines on the plot's rhs.

Although parameter $\Omega_{G0}$ cannot be bounded by observational data on the Hubble factor,
it is related to $a_{in}$, and to the coincidence problem inherent to the IBEG model when $a_{in}\sim 1$ \cite{Luc18}. Given the observational and theoretical bounds on the rest of the parameters (specially  $x$, which is found to be close to unity), no fine tuning on $\Omega_{G0}$ is needed to avoid $a_{in}\sim 1$. On the contrary, only a small range of values $\Omega_{G0}$ close to zero leads to $a_{in}\sim 1$.

The IBEG-model parameter $\Omega_{m0}$ should not be compared with the $\Lambda$CDM-model DM parameter, as   $\Omega_{m0}$ is only a fraction of the DM energy density  $\Omega_{dm0}$ while other components have a negative contribution due to the coupling term; as   $\Omega_{G0}$ cannot be bounded by the observational data, we neither can give observational bounds on $\Omega_{dm0}$ in the IBEG model,
which makes it useless to compare it with the $\Lambda$CDM model.

\subsection{PGW late evolution and numerical results}

The   PGW amplitude, $\mu(\eta)$, evolves with conformal time according to eq.  (\ref{eqmu}). The    FLRW universe dynamics affects the amplitude evolution through the potential term $a''/a$, which relates  to the Hubble-factor   variable
 as $a d\eta= dt=da/(H a)$. The potential $a''/a$ can be expressed in terms of $a$ as
\be
\frac{a''(a)}{a}=2a^2 H^2(a)+a^3 H(a) \frac{dH(a)}{da},
\label{numpot}
\ee
while eq. (\ref{eqmu}) is transformed to
\ben
a^4 H^2(a) \frac{d^2 \mu(a)}{da^2}  &+&\left[ 2a^3 H^2(a)+a^4 H(a) \frac{dH (a)}{da}  \right ]\frac{d \mu (a)}{da} \nonumber\\&+&\left [k^2-2a^2 H^2(a)+a^3 H(a) \frac{dH (a) }{da} \right ]\mu(a)=0.
\label{nummu''}
\een

We note that while integrating eq. (\ref{eqmu}) in terms of the conformal time $\eta$, $a''/a$ is the only term present, as an additional term proportional to $d \mu/da$ appears when integrating in terms of $a$. For the late-evolution IBEG model,  and  Hubble factor $H(a)$   given in eq. \ref{H},
 we can compute $dH(a)/da$, and  eventually  solve equation (\ref{eqmu}) by numerical methods for different  free-parameter sets.

We consider the  free-parameter set of $\Omega_{m0}$, $\Omega_{i0}$, $\Omega_{G0}$, and $x$, while we fix $\Omega_{b0}H_0^2=0.022\ ({\rm km\, s^{-1}\, Mpc^{-1}})^2$ in order to
compute the instant for which the creation process starts $a=a_{in}$. It is possible to divide eq. (\ref{nummu''}) by $H_0^2$ and to set $H_0=1$ at this point, defining the scale of frequencies of the PGW through wave number $k$. We set $a_2=.0001$ as the beginning of the dust era.  If $a_{in}\leq 0.0001$, we use the Hubble factor as eq. (\ref{H}) for $a\in[0.0001,1]$ in order to solve numerically eq. (\ref{nummu''}) for different $k$ choices. We use initial conditions at  instant $a=0.0001$ as  $\mu(a=0.0001)=\mu_R(a=0.0001)$ and $\frac{d \mu}{da}(a=0.0001)=\frac{d\mu_R}{da}(a=0.0001)$.

On the other hand, if $a_{in}>0.0001$, we first solve eq. (\ref{nummu''}) with the Hubble expansion rate dominated by non-relativistic matter (a mixture of baryonic matter and CDM) as $H(a)=H_{in}(a_{in}/a)^{3/2}$ (where $H_{in}$ is the Hubble factor in eq. (\ref{H}) evaluated at scale factor $a_{in}$) for $a\in[0.0001,a_{in}]$
to obtain a first solution $\mu_1$ (with initial conditions  $\mu_1(a=0.0001) $ and $\frac{d \mu_1}{da}(a=0.0001) $). Then, we solve  eq. (\ref{nummu''})
with the  Hubble factor in eq. (\ref{H}) for $a\in[a_{in},1]$ to obtain a second solution $\mu_2(a)$ with initial conditions at $a=a_{in}$, with  matching of the first solution $\mu_1$ at $a=a_{in}$ ($\mu_2(a=a_{in})=\mu_1(a=a_{in})$ and $\frac{d \mu_2}{da}(a=a_{in})=\frac{d \mu_1}{da}(a=a_{in})$).

\begin{figure}
\centering
\includegraphics[scale=0.3]{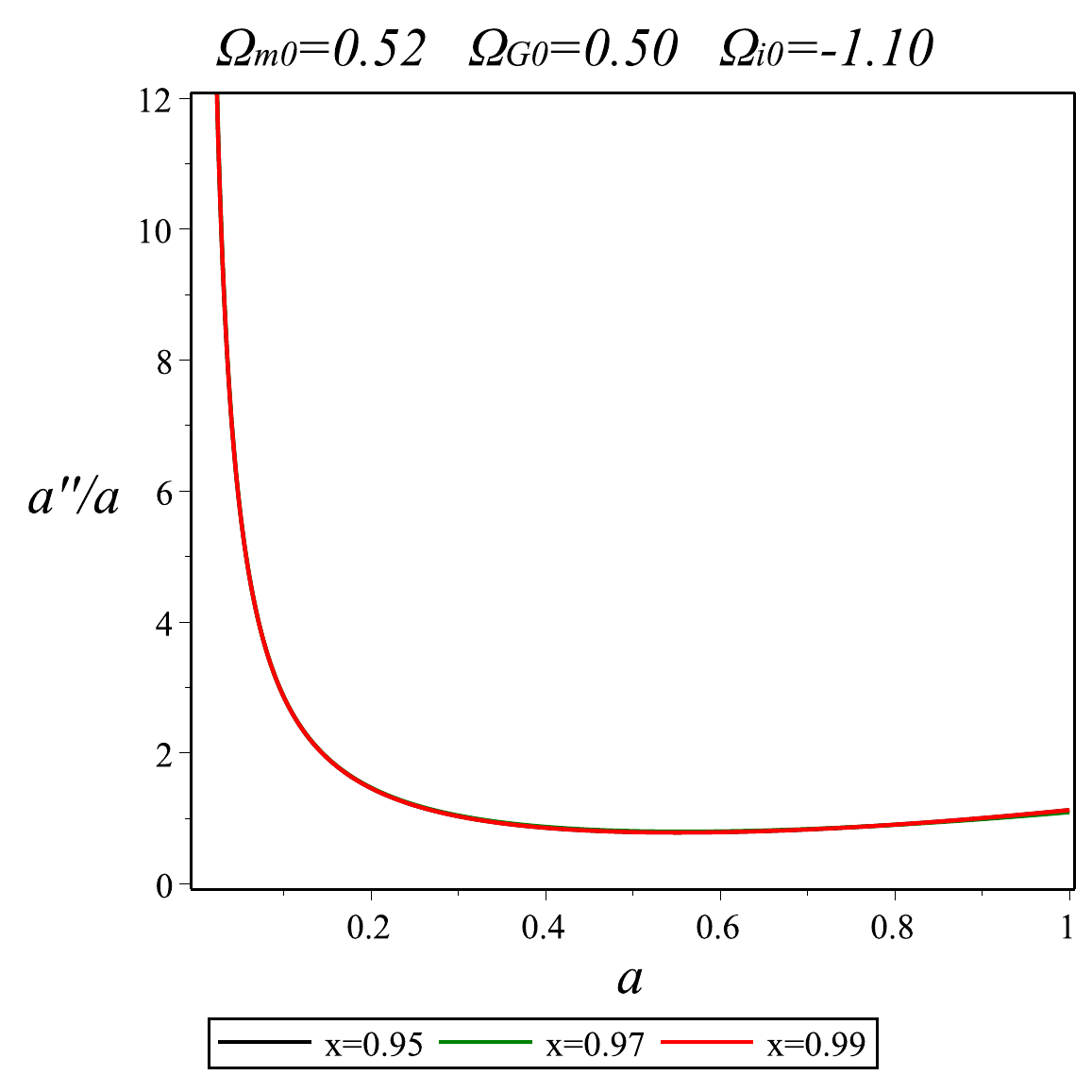}
\includegraphics[scale=0.3]{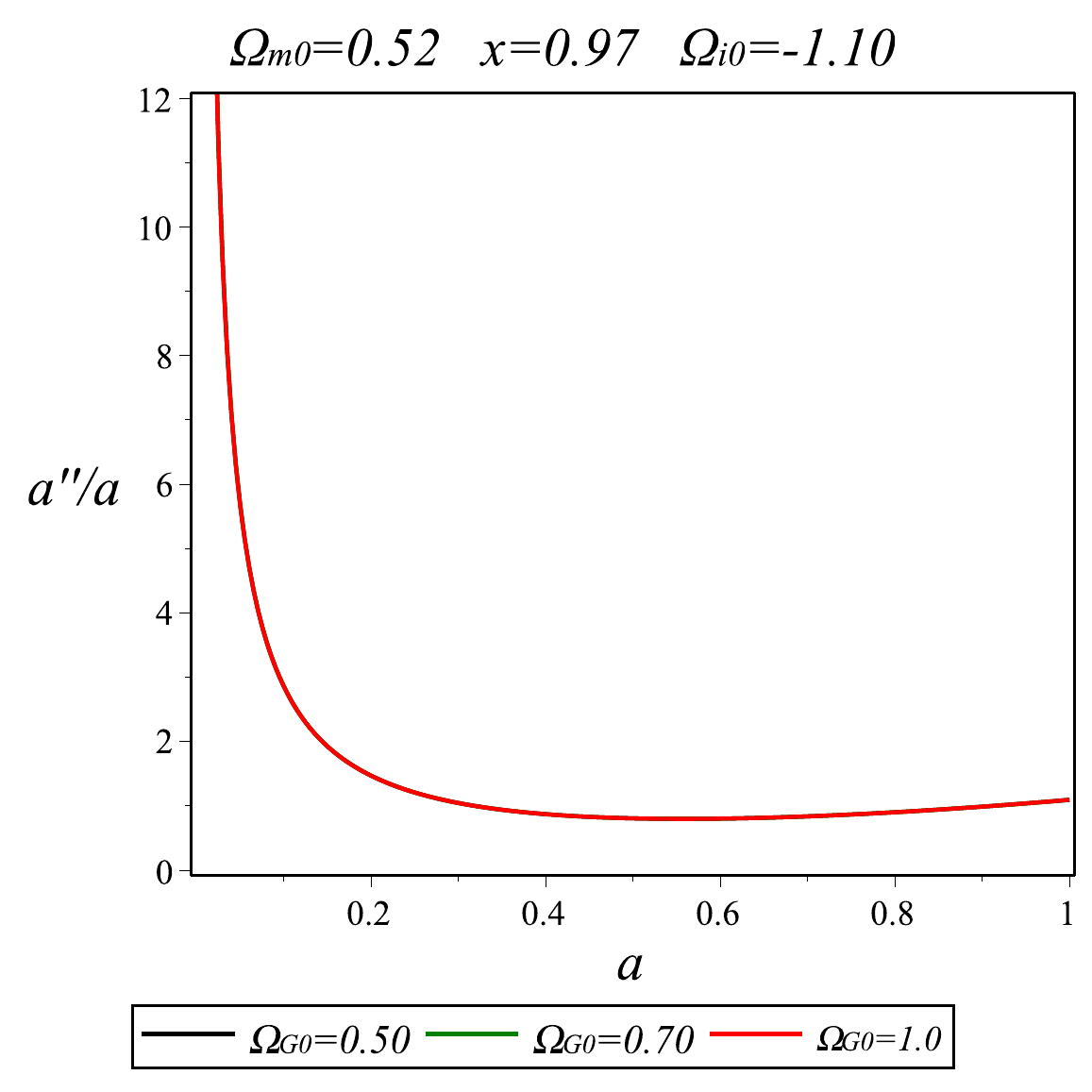}
\includegraphics[scale=0.3]{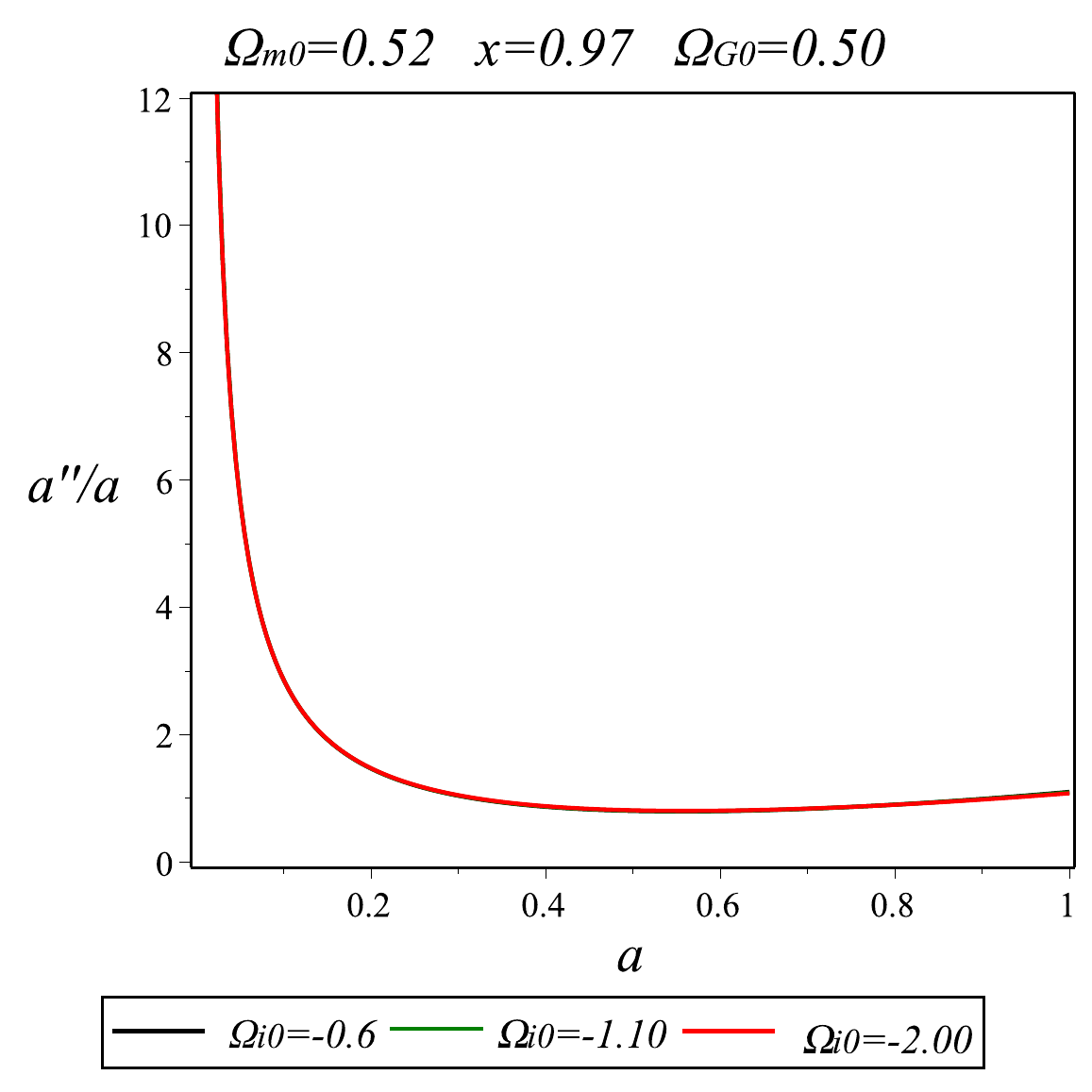}
\includegraphics[scale=0.3]{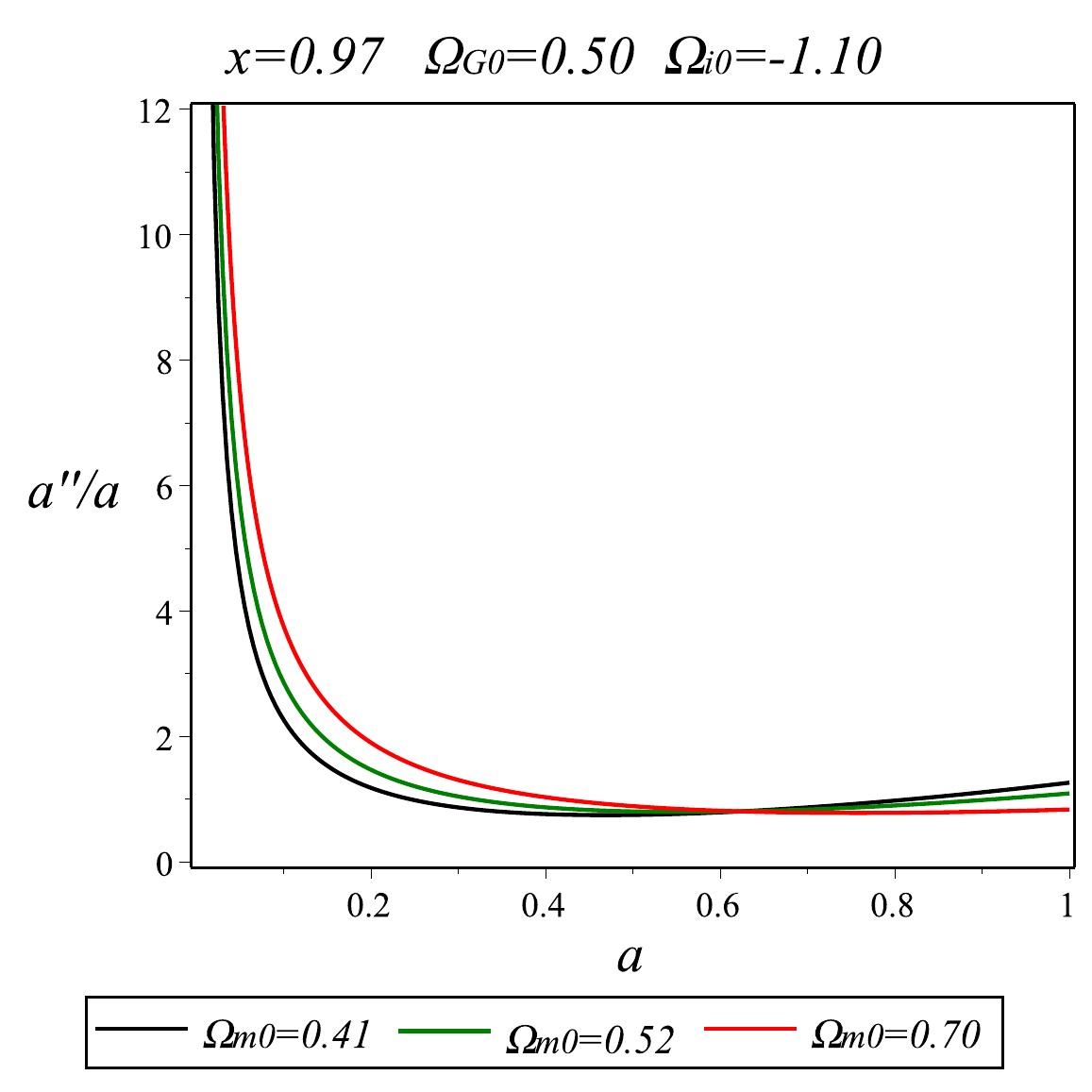}
\caption{ Potential $a''/a$ vs. scale factor $a$, with each plot variating $x$, $\Omega_{  G0}$, $\Omega_{  i0}$  and $\Omega_{m0}$, and fixing the other parameters. The one parameter with a noticeable impact on the potential is $\Omega_{m0}$, as  the rest of     the free
parameters   leave the potential unchanged up to eye view.}
\label{fig2}
\end{figure}

We first compute the potential term in  eq. (\ref{numpot}) for different   free-parameter   choices  in order to determine which one has the biggest impact on the amplitude. We note that   $\Omega_{G0}$ affects the amplitude of PGW only via  $a_{in}$, as the Hubble factor (and, consequently, eq. (\ref{nummu''})) do not explicitly depends on $\Omega_{G0}$. The parameter $\Omega_{m0}$ is chosen in the
2-$\sigma$ region shown in figure \ref{fig1}, while $\Omega_{i0}$ is chosen to lie on the $\Omega_{dm0}+\Omega_{G0}>0$ region. Figure \ref{fig2} shows the potential vs. scale factor for different   parameter choices. The one parameter with a noticeable impact on the potential is $\Omega_{m0}$, while the rest of     the free
parameters   leave the potential unchanged up to eye view.

\begin{figure}
\centering
\includegraphics[scale=0.3]{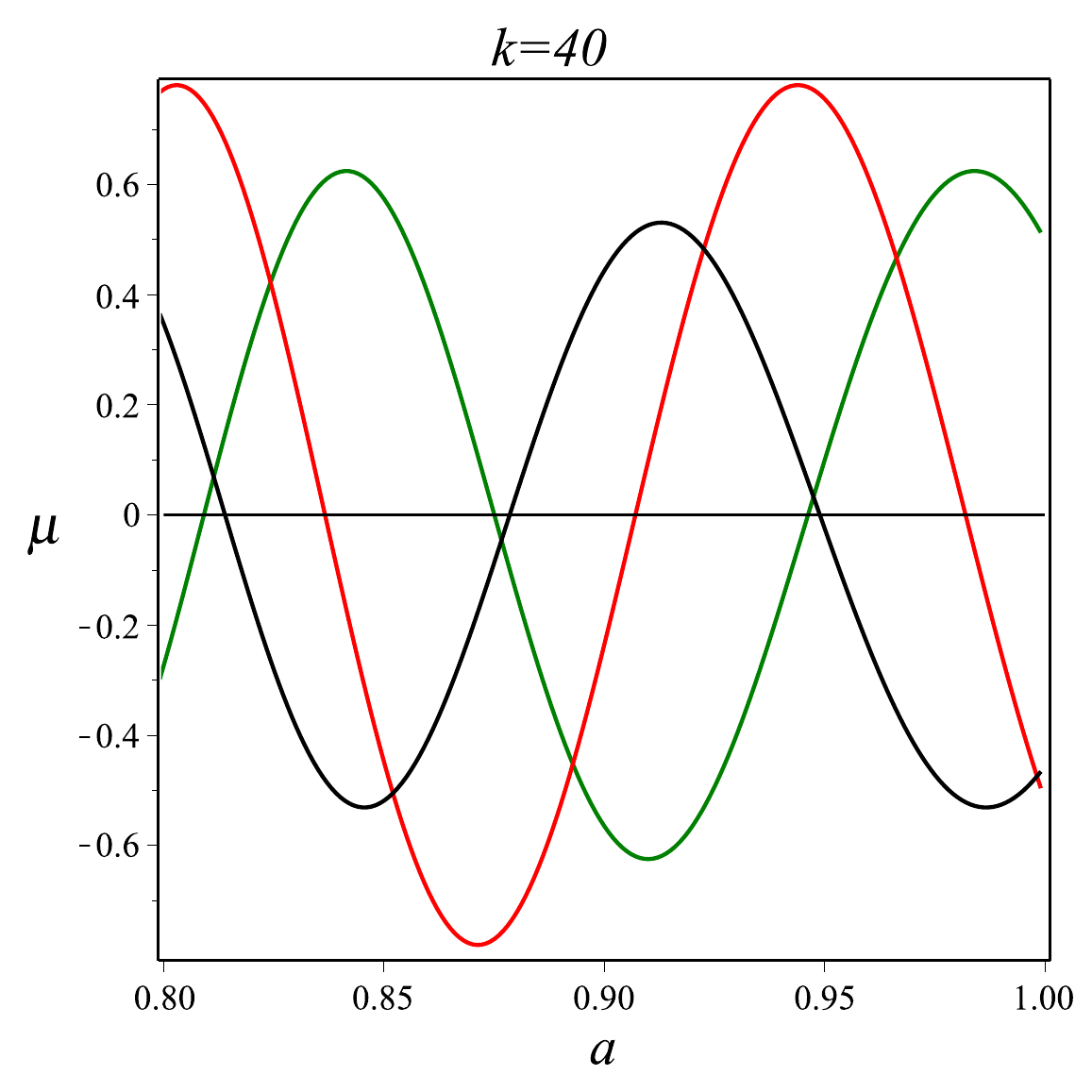}
\includegraphics[scale=0.3]{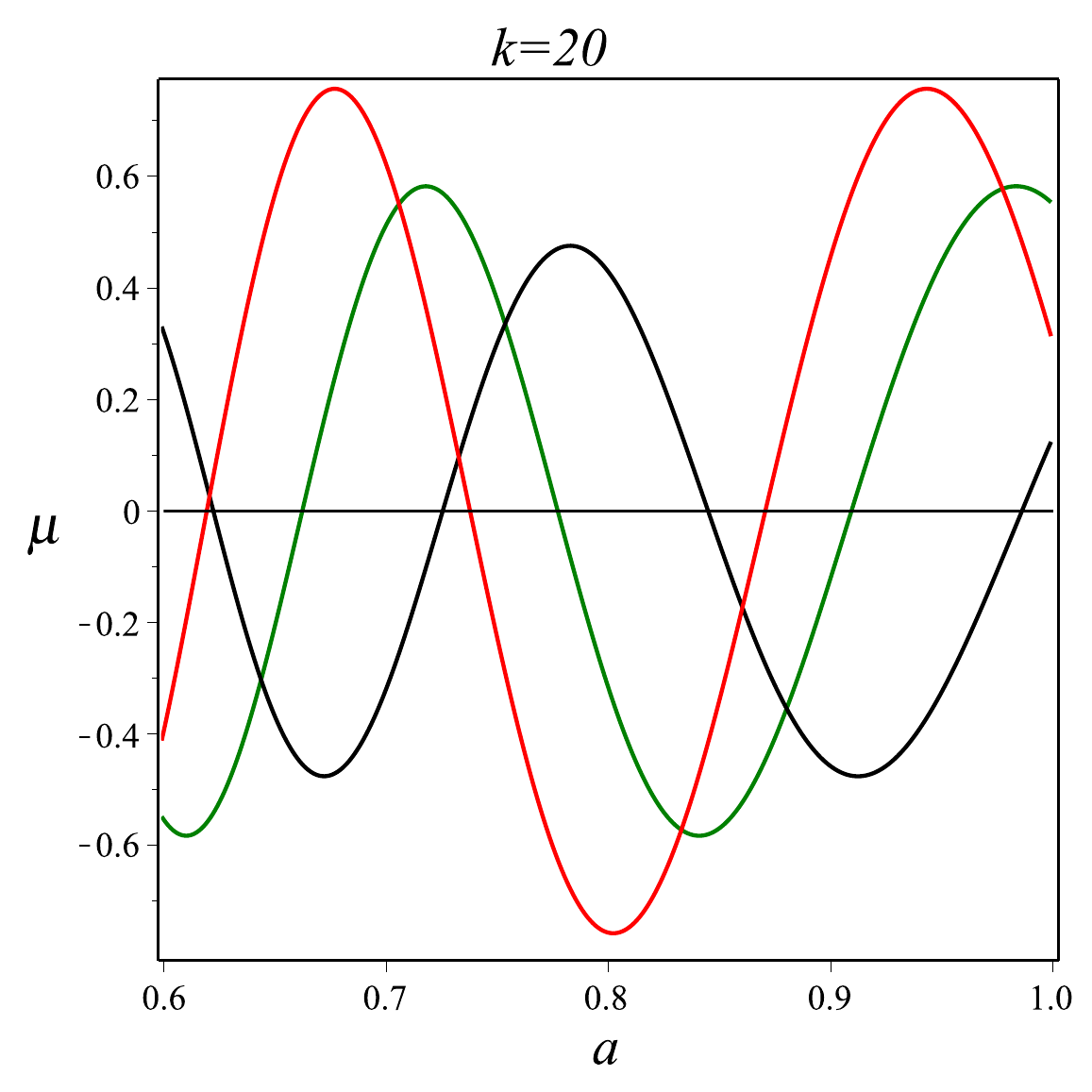}
\includegraphics[scale=0.3]{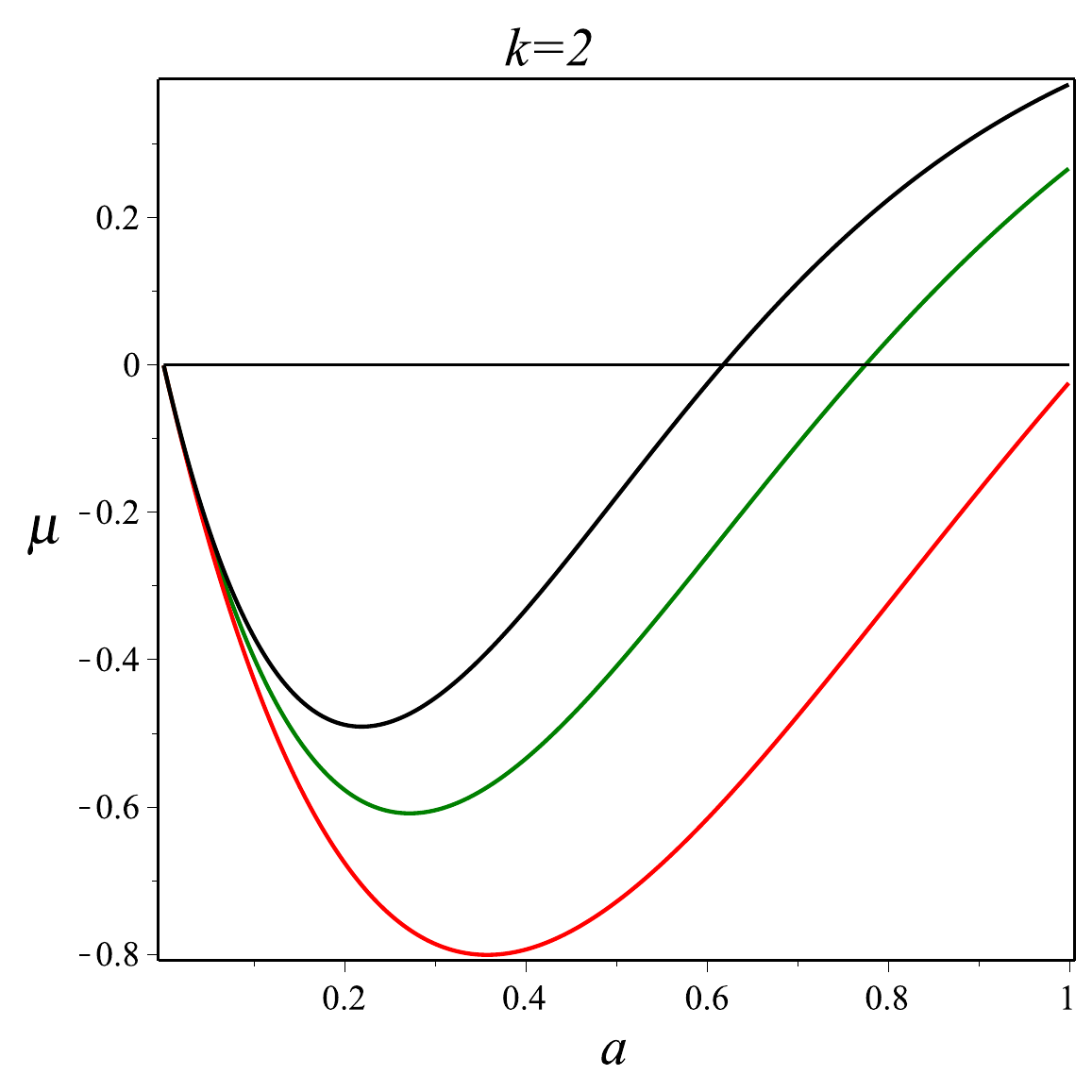}

\caption{Evolution of PGW amplitude for the IBEG model from the beginning of dust $a=0.0001$ until today $a=1$ era with $x=0.97$, $\Omega_{G0}=0.5$, and $\Omega_{i0}=-1.10$ for different wave number choices: $k=40$, $k=20$, and $k=2$. The black line represents the evolution with $\Omega_{m0}=0.41$, the green line represents $\Omega_{m0}=0.52$, and the red line represents $\Omega_{m0}=0.70$}
\label{fig3}
\end{figure}

The  PGW  amplitude  depends on  $\Omega_{m0}$ as expected (figure \ref{fig3}). The larger   parameter $\Omega_{m0}$ the larger the resulting PGW amplitude. Also,
for different    wave numbers $k$, the amplification of the same free parameters   varies as well. In the next section, we compute the  PGW power spectrum, related to the
amplitude, as a function of  $\Omega_{m0}$.

\section{Power spectrum}
\label{s4}
In this section, we consider the best-fit value for the  Hubble factor $H_0=70\  {\rm km}/( {\rm Mpc\  s})\simeq 2.27 \times 10^{-18}\ {\rm s}^{-1}$ obtained in \cite{Luc18}, as we use  units for the frequency and wave number.
  The wave number $k$ is not a physical quantity (it is defined a comoving
  quantity), while the corresponding physical frequency is defined as $\omega (a) =k/a$. Given that $a_0=1$, we use that the frequency of the PGW observed today corresponds to the wavenumber $k=\omega$.
 The    PGW amplitude depends on their wave number and the amplification
  regime experienced through the expansion of the universe.

We   assume a typical slow-roll de Sitter inflation with $H_1= 10^{35}\ {\rm s}^{-1}$, at the inflation-radiation transition $\eta_1$.
  It is straightforward that
\[
a_1=0.0001\left(H(a_2)/H_1\right)^{1/2},
\]
with $H(a_2)$ being the Hubble factor at the beginning of the dust/IBEG era, which depends on the IBEG-model free parameters   as well. In all cases, the bound is of order ${K_1}^2=a_1 H_1\sim 10^{11}{\rm s}^{-1}$. As stated in section \ref{s2}, waves with $k\gg K_1\sim 10^{11}\ {\rm s} ^{-1}$ did not experience any adiabatic amplification and have an amplitude several orders of magnitude smaller at present than at the instant they were generated. Consequently, we can assume that those waves do not contribute to the PGW power spectrum.

PGW with wave number $k^{2}\ll K_1$ but $k^{2}\gg\frac{a^{\prime\prime }}{a}(a=a_2)$ evolve as free waves, after a first amplification regime during inflation, and are not affected by the   late-universe dynamics. For $a=a_2=0.0001$, we define this bound as ${K_2}^2=\frac{a''}{a}(a_2)$, which also depends on free parameters through $H(a_2)$ and $\frac{dH(a_2)}{da}$. The  PGW amplitude in this regime is proportional to

\[
\frac{C_{D}}{C_{I}}\sim \left\{
\begin{array}{c}
1\qquad (k\gg K_1), \\
k^{-2}\qquad (K_1\gg k\gg K_2), \\
\end{array}%
\right.
\]
and they are considered in the power spectrum.

PGW with $k \ll K_2$, undergo an amplification during inflation and a second one in the dust/IBEG eras. On the other hand, the perturbations whose wave number is $k\ll (a^{\prime \prime }/a)(a=1)=K_0^2$ have wavelengths larger than the Hubble radius of the universe, i.e. on the whole history of the universe, they have not completed a single
period of oscillation. Those perturbations cannot be considered as physical waves. This puts a lower bound on the wave number for the PGW spectrum, $K_0$, that also depends on the free parameters.

The waves with $K_2 \ll k \ll K_1$ have the same power spectrum as in \cite{Almazan14}
\be
P(k)=\frac{\hbar }{4\pi ^{2}c^{3}}a_{1}^{4}H_{1}^{4}k ^{-1},
\ee
where the initial perturbations on the inflationary field are   Gaussian, thus, $C_I \propto k^2$ \cite{grish74}.

The power spectrum of the  waves with $K_0 \ll k \ll K_2$ is obtained by numerically computing $\mu(a)$ for each $k$ as in the previous section. We consider, then,
\be
P(k)=\frac{\hbar }{4\pi ^{2}c^{3}}k^3 |\mu_{rms}|^2
\ee
where $|\mu_{rms}|$ is the root mean square of the PGW as
\be
|\mu_{rms}|^2=\frac{2}{(1-a_p)}\int^1_{a_p}\mu(a)^2 da
\ee
where $a_p$ corresponds to the scale factor for which the corresponding $k\eta(a_p)=2\pi$, i.e., the scale factor at the start of the last oscillation of the PGW.

\begin{figure}
\centering
\includegraphics[scale=0.5]{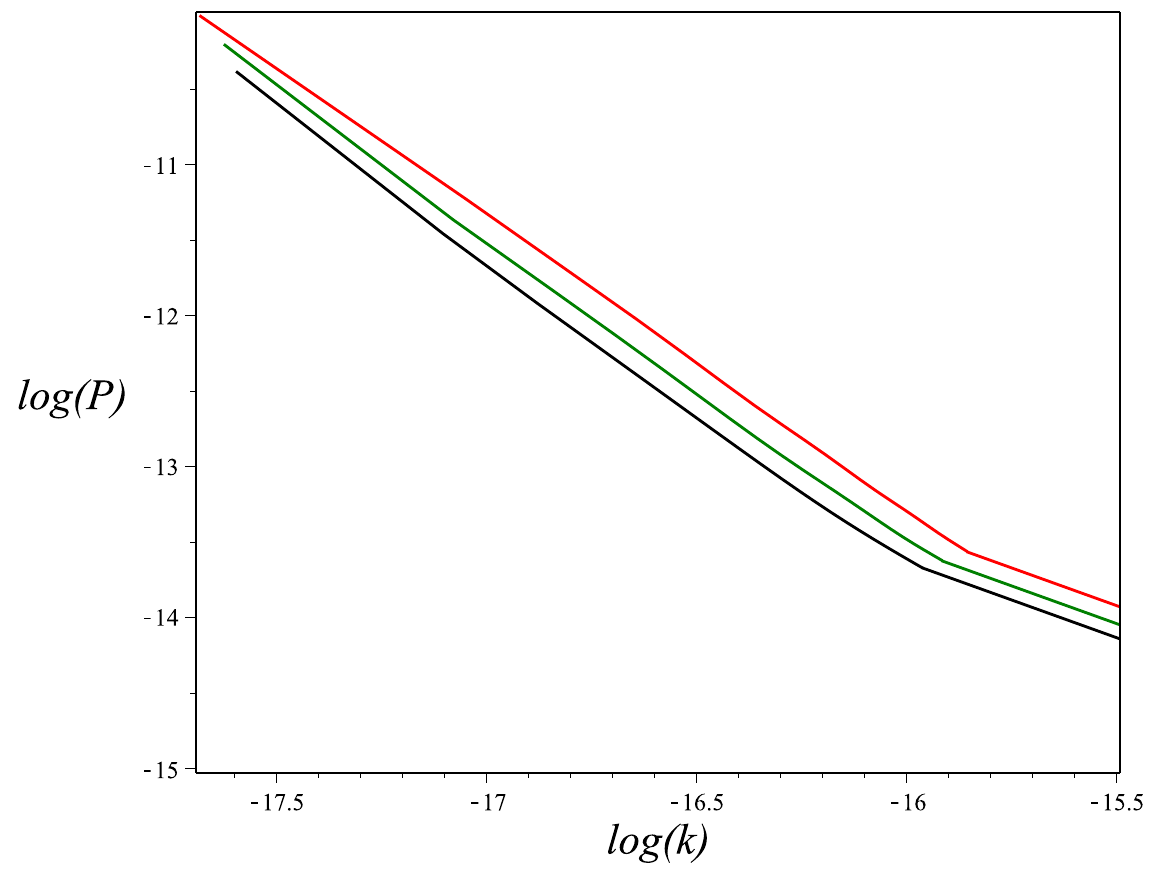}

\caption{ IBEG-model power spectrum  $P$ with $x=0.97$, $\Omega_{G0}=0.5$, and $\Omega_{i0}=-1.10$. The black line represents $\Omega_{m0}=0.41$, the green line  $\Omega_{m0}=0.52$, and the red line  $\Omega_{m0}=0.70$. $P$ and the frequency $k$ are expressed in ${\rm erg}  \ {\rm s/cm^3}$ and ${\rm s}^{-1}$, respectively.}
\label{fig4}
\end{figure}

Figure \ref{fig4} shows ${\rm log}_{10}(P(k))$ vs ${\rm log}_{10}(k)$ for the    IBEG-model PGW, for $x=0.97$,
$\Omega_{G0}=0.5$, $\Omega_{i0}=-1.10$, and   $\Omega_{m0}$ choices. The power spectrum dependence on   $\Omega_{m0}$ is expressed in the slope of the
 line for $k<K_2\approx10^{-16}$ and also on the bounds $K_1$, $K_2$, $K_0$. The rest of free parameters do not affect significatively the power-spectrum slope  or   $K_1$, $K_2$, $K_0$.

\begin{figure}
\includegraphics[scale=0.6]{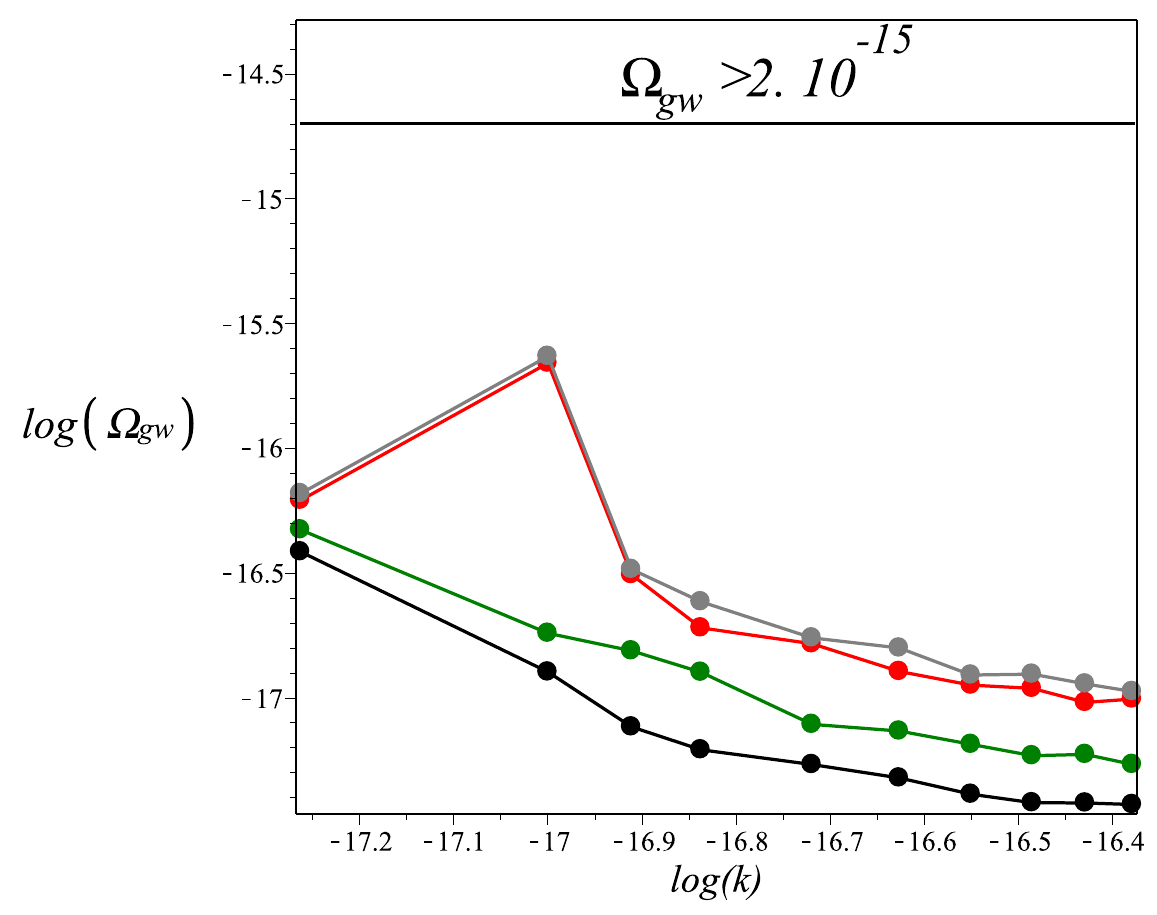}
\caption{$\log(\Omega_{gw}(a=1))$ computed at different discrete choices of wave number ($k\sim10^{-17}$) for $H_1=10^{35}\ {\rm s}^{-1}$, $x=0.97$, $\Omega_{G0}=0.5$, and $\Omega_{i0}=-1.10$, Hubble parameter  $h=0.7$, and different values of $\Omega_{m0}$ parameter (the black line represents $\Omega_{m0}=0.41$, the green   $\Omega_{m0}=0.52$, the red    $\Omega_{m0}=0.70$, and the grey   $\Omega_{m0}=0.75$). The top straight line separates the region fulfilling the observational bound $\Omega_{gw}(a=1)<2 \cdot 10^{-15}$ from the region excluded by observations.}
\label{fig5}
\end{figure}

Another relevant definition is the fraction of energy density per frequency \cite{watanabe06}
\be
\Omega_{gw}(k,a)= \frac{3}{8 \pi G H^2} \frac{d \rho_{gw}}{d(\ln k)}= \frac{4}{3 \pi a^2 H^2}\frac{H_{1}}{M_{pl}^2}|T'|^2,
\ee
where $T'$ is a transfer function  related to $(\mu/a)'$ for waves with $k<K_2$ during the IBEG stage of expansion, and $\rho_g$ is the energy  density of the PGW obtained from tensor first order perturbation theory as

\be
\rho_{gw}= \frac{1}{32 \pi G a^2} \langle {h_{ij}}' {h^{ij}}' \rangle.
\ee
The observational bounds suggest that present day ($a=1$) $\Omega_{gw} h^2 < 10^{-15}$ for frequencies $k\sim10^{-17}$ with $h=H_0/(100\ {\rm km}/( {\rm Mpc\  s}))$ \cite{guzzetti16}. Although these bounds are strongly related to inflation and reheating parameters of the cosmological model, the late-accelerated expansion model should be taken into account, as  for low frequencies $|(\mu/a)'|^2$
depends on $H$. In our case, assuming the above  free-parameter  choices ($x=0.97$, $\Omega_{G0}=0.5$, and $\Omega_{i0}=-1.10$)  with $h=0.7$, present-day $\Omega_{gw}$ at several wave numbers of order  $k=10^{-17}$
were computed for some values of $\Omega_{m0}$: $\Omega_{m0}=0.41$, $\Omega_{m0}=0.52$, $\Omega_{m0}=0.70$,and  $\Omega_{m0}=0.75$. The results are shown in figure \ref{fig5}. The last $\Omega_{m0}$ value is out of the 2-$\sigma$ region of figure \ref{fig1}, but close enough  as not to be excluded beforehand.

We note that the present-day $\Omega_{gw}$ bound is heavily dependent on the inflation model   considered, and it is not a PGW prediction per se, but an order of magnitude estimate. It would be erroneous   to strictly bound  parameters of the late-acceleration model considered (IBEG model in this work) from it. However, it is safe to conclude that the late accelerated-model parameters have a non-negligible impact on $\Omega_{gw}$ as do inflation parameters (we have used $H_1=10^{35}\ {\rm s}^{-1}$). In our model,  only parameter $\Omega_{m0}$ has a noticeable impact on  present-day $\Omega_{gw}$, and, the higher  $\Omega_{m0}$ is, the higher $\Omega_{gw}$. At wave number $k=10^{-17}$, $\Omega_{gw}$ reaches the highest value for  $\Omega_{m0}=0.75$ (with $\Omega_{gw}$  for $\Omega_{m0}=0.70$ slightly lower), but an order of magnitude smaller than the observational limit. Choosing $H_1=10^{36}\ {\rm s}^{-1}$ would lead to $\Omega_{gw}$  for both $\Omega_{m0}=0.75$ and $\Omega_{m0}=0.70$ reaching the observational limit, while the smaller $\Omega_{m0}$ values would be still one or two order of magnitude under it.

\section{Conclusions}
\label{s6}

PGWs are second-order tensorial  wave-like solutions to the cosmological Einstein equations that are generated and amplified by the universe dynamics. Inflation and early phase transitions are crucial to the amplification process, but the late accelerated expansion has also an
important contribution to the low-frequency wave  evolution \cite{iz04,Almazan14}.

In particular, the IBEG model is applied for a flat late-acceleration expansion    with a detailed microscopical description. The IBEG model has four free parameters:  $\Omega_{G0}$,  $\Omega_{m0}$, related to the dark-energy, rest-mass energy density and the dark-matter term scaling as a mass term, respectively;  $\Omega_{i0}$,  the self-interaction intensity; $x$,   the   energy exchange rate. Other parameters of the model are the Hubble constant $H_0$ and the baryonic matter parameter $\Omega_{b0}$. The free parameters can be bounded by observational data \cite{Luc18}.

The PGW amplitude evolution in the IBEG model depends on the free parameters through the Hubble factor $H(a)$ in eq. \ref{H} and its derivative $dH/da$. When considering different values for the free parameters we conclude that only parameter $\Omega_{m0}$ has a noticeable impact. The higher    $\Omega_{m0}$,  the larger the PGW amplitude for constant wave number. The PGW power spectrum depends consequently on the parameter $\Omega_{m0}$. Additionally, the fraction of energy density per frequency $\Omega_{gw}(k,a)$ has a non trivial dependency on parameter $\Omega_{m0}$ at low frequencies.
We also derived  the model's   PGW power, which is consistent  with observational bounds from below, similarly to other models.

{\bf Acknowledgements}

 The
authors  acknowledge support from  the Direcci\'on General de Asuntos del Personal Acad\'emico, at the Universidad Nacional Aut\'onoma de M\'exico,  through  Project IN117020.
{\bf Data access statement}

The authors declare that the data supporting the findings of this study are available within the article.


\begin{thebibliography}{99}
\section*{References}
\bibitem{Einstein1916} A. Einstein, Preuss. Akad. Wiss. Berlin, Sitzber, 688 (1916).
\bibitem{Hulse75} R. A. Hulse and J. H. Taylor, Astrophys. J. Lett. 195, L51 (1975).
\bibitem{ligo16} B. P. Abbott et al. [LIGO Scientific and Virgo Collaborations], Phys. Rev. Lett. 116, 061102 (2016).

\bibitem{schutz86} Schutz, B. F., Nature, 323, 310 (1986).

\bibitem{ligo17}  Abbott, T. D. {\em et~al.\/}, The LIGO Scientific Collaboration {\em Nature\/} {\bf 551} 85-88 (2017).

\bibitem{ade2016planck}
Ade P~A, Aghanim N, Arnaud M, Ashdown M, Aumont J, Baccigalupi C, Banday A,
  Barreiro R, Bartlett J, Bartolo N {\em et~al.\/} {\em Astronomy \&
  Astrophysics\/} {\bf 594} A13 (2016).

\bibitem{Riess20162}
Riess A~G, Macri L~M, Hoffmann S~L, Scolnic D, Casertano S, Filippenko A~V,
  Tucker B~E, Reid M~J, Jones D~O, Silverman J~M {\em et~al.\/} {\em The
  Astrophysical Journal\/} {\bf 826} 56 (2016).

\bibitem{Lifs} E. Lifshitz, J. Phys. USSR \textbf{10}, 116 (1946).

\bibitem{grish74} L. P. Grishchuk, Zh. Eksp. Teor. Fiz. \textbf{67}, 825 (1975) [Sov. Phys. JETP 40, 409 (1975)].
\bibitem{stochLIGO17} B. P. Abbott et al. [LIGO Scientific and Virgo Collaborations], 	Phys. Rev. Lett. 118, 121101 (2017).
\bibitem{eLISA} https://www.elisascience.org/

\bibitem{sel}
U. Seljak and M. Zaldarriaga, Physical Review Letters, \textbf{78}, 2054 (1997).
D. N. Spergel and M. Zaldarriaga, Physical Review Letters, \textbf{79}, 2180 (1997)

\bibitem{cheng} Cheng Cheng, Qing-Guo Huang, arXiv:1403.5463.


\bibitem{copeland2006dynamics}
Copeland E~J, Sami M and Tsujikawa S 2006 {\em International Journal of Modern
  Physics D\/} {\bf 15} 1753-1935


\bibitem{Wang2016dmdeinteract}
Wang B, Abdalla E, Atrio-Barandela F and Pav\'on D 2016 {\em Reports on Progress
  in Physics\/} {\bf 79} 096901

\bibitem{barrow2006} J. D. Barrow  and T. Clifton,    Phys. Rev. D {\bf 73} 103520 (2006).
\bibitem{Shahalam15} M. Shahalam et al., {\em Eur. Phys. J. C } {\bf 75} 395 (2015).

\bibitem{SRay11} S.  Ray et al., {\em Int. J .Theor. Phys.} {\bf 50}  939-951, (2011).

\bibitem{Dymnikova00} I. G. Dymnikova and  M. Yu. Khlopov,
{\em Mod. Phys.
Lett. A }  {\bf 15},  2305-2314 (2000).

\bibitem{Dymnikova01} I. G. Dymnikova  and  M. Yu. Khlopov,
{\em Eur. Phys. J. C}
  V. 20, PP. 139-146 (2001).

\bibitem{iz04} G. Izquierdo and D. Pav\'on, {\em Physical Review D} \textbf{70} 084034 (2004).






\bibitem{sasaki18} T. Sasaki and H. Suzuki, Progress of Theoretical and Experimental Physics, Volume, 11, 113E03 (2018).
\bibitem{Almazan14} M. L. Sosa-Almazan and G. Izquierdo, General Relativity and Gravitation \textbf{46} 1759 (2014).

\bibitem{iz10} G. Izquierdo and J. Besprosvany, Classical and Quantum Gravity, {\bf 27} 065012 (2010).

\bibitem{bespro2015} J. Besprosvany and G. Izquierdo, Classical and Quantum Gravity, {\bf 32} 055015 (2015).


\bibitem{Luc18} H. E. Lucatero-Villase\~nor, G. Izquierdo and J. Besprosvany, General Relativity and Gravitation, \textbf{50}, 151 (2018).

\bibitem{grish93} L. P. Grishchuk, Class. Quantum Grav. \textbf{10}, 2449
(1993).


\bibitem{Khlopov85} M. Yu. Khlopov, B. A. Malomed and Ya. B. Zeldovich,{\em Mon. Not. Roy. astr. Soc.}  {\bf 215},  575-589 (1985).




\bibitem{watanabe06} Y. Watanabe and E. Komatsu, Physical Review D \textbf{73} 123515 (2006).

\bibitem{guzzetti16} M. C. Guzzetti, N. Bartolo, M. Liguori, and S. Matarrese, Rivista del Nuovo Cimento, Vol. 39, Issue 9, 399-495 (2016).

\end{thebibliography}
\end{document}